\documentclass[article,shortnames, nojes]{jes}

\usepackage{graphics}
\usepackage{graphicx}
\usepackage{amsmath}
\usepackage{amssymb}
\usepackage{accents}
\usepackage{fancyhdr,lastpage}
\usepackage{setspace}  
\usepackage[small, bf]{caption} 
\usepackage{booktabs}
\usepackage{bbold}
\usepackage{multirow}
\usepackage{lscape}
\usepackage{rotating}
\usepackage{caption}
\usepackage{subcaption}
\usepackage{bigints}                  
\usepackage{hyphenat}             
\usepackage{ragged2e}
\usepackage{adjustbox}             
\usepackage{natbib}

\usepackage{ae}
\usepackage{color}
\usepackage{url}
\usepackage{lineno}

\newcommand{\notwo}{NO$_2$ }

\newcommand{\acid}{Acid/Aerosol }
\newcommand{\acidns}{Acid/Aerosol}
\newcommand{\s}{\mathbf{s}}

\defcitealias{epano2}{US EPA, 2010}
\defcitealias{epadata}{US EPA, 2011}
\defcitealias{epa2004}{US EPA, 2004}

\defcitealias{cmaq2000}{US EPA, 2014}
\defcitealias{byun2006}{Byun and Schere, 2006}
\defcitealias{hogrefe2009}{Hogrefe et al., 2009}

\defcitealias{ctdot2000}{CT DOT, 2001}
\defcitealias{madot}{MA DOT, 2012}
\defcitealias{nydot}{NY DOT, 2012}

\author{Owais Gilani\\ Bucknell University \\ Lewisburg, PA, USA \And 
        Lisa A. McKay\\ Yale University \\ New Haven, CT, USA \And 
	Timothy G. Gregoire \\  Yale University \\ New Haven, CT, USA \AND
	Yongtao Guan \\  University of Miami \\ Miami, FL, USA \And
	Brian P. Leaderer \\  Yale University \\ New Haven, CT, USA \And
	Theodore R. Holford \\  Yale University \\ New Haven, CT, USA}

\title{Spatiotemporal Calibration of Atmospheric Nitrogen Dioxide Concentration Estimates From an Air Quality Model for Connecticut} 

\Plainauthor{O. Gilani, L. McKay, T. Gregoire, Y. Guan, B. Leaderer, T. Holford} 
\Plaintitle{Spatiotemporal Calibration of Atmospheric Nitrogen Dioxide Concentration Estimates From an Air Quality Model for Connecticut} 
\Shorttitle{Spatiotemporal Calibration of CMAQ NO$_2$} 

\Abstract{A spatiotemporal calibration and resolution refinement model was fitted to calibrate nitrogen dioxide (NO$_2$) concentration estimates from the Community Multiscale Air Quality (CMAQ) model, using two sources of observed data on NO$_2$ that differed in their spatial and temporal resolutions. To refine the spatial resolution of the CMAQ model estimates, we leveraged information using additional local covariates including total traffic volume within 2 km, population density, elevation, and land use characteristics. Predictions from this model greatly improved the bias in the CMAQ estimates, as observed by the much lower mean squared error (MSE) at the NO$_2$ monitor sites. The final model was used to predict the daily concentration of ambient \notwo over the entire state of Connecticut on a grid with pixels of size 300 x 300 m. A comparison of the prediction map with a similar map for the CMAQ estimates showed marked improvement in the spatial resolution. The effect of local covariates was evident in the finer spatial resolution map, where the contribution of traffic on major highways to ambient \notwo concentration stands out. An animation was also provided to show the change in the concentration of ambient \notwo over space and time for 1994 and 1995.
}
\Plainkeywords{SCARR model,  CMAQ, resolution refinement, integrated exposure modeling, ambient air pollution, Kalman filter} 

\Address{
  Owais Gilani\\
  Department of Mathematics, Bucknell University\\
  Lewisburg, PA 17837\\
  Telephone: +1/570-577-1391\\
  E-mail: \email{owais.gilani@bucknell.edu}
}

\begin{document}

\setkeys{Gin}{width=1.0\textwidth}  


\section[Introduction]{Introduction} \label{Paper2intro} 

Nitrogen dioxide (NO$_2$) is a highly reactive gas that contributes to the formation of ground-level ozone and fine particle pollution, and is believed to be associated with adverse respiratory health effects \citep{ISA2008}. A detailed analysis of the effects of atmospheric pollutants such as NO$_2$ on various health outcomes requires access to data on the concentration of the pollutant on a fine spatial and temporal scale, which is rarely available. However, we often have data on the concentration of atmospheric pollutants for a given region and time period from different sources that differ in their spatial and temporal resolutions, as well as in their measurement accuracy. 

Fixed site air quality monitoring stations, such as the US Environmental Protection Agency's (EPA) monitoring stations \citepalias{epadata}, record pollutant concentration data on a dense temporal scale (hourly), but the network of monitoring sites is generally spatially very sparse (e.g. only four sites in Connecticut (CT)), which doesn't allow for accurate modeling at sites far away from the monitoring sites. On the other hand, data collected at many different spatial locations using passive sampling as part of environmental epidimiologic studies, such as the \acid study \citep{triche2002}, generally provide an aggregate measure of the pollutant concentration over relatively long time periods (1-2 weeks), resulting in spatially dense but temporally sparse data. Such data sources do not allow accurate estimation of pollutant concentrations at a fine temporal scale. 

In the absence of a single source of observed data on the concentration of a pollutant that is both spatially and temporally dense, deterministic meteorological air quality models, such as the Community Multiscale Air Quality (CMAQ) model \citepalias{byun2006, hogrefe2009, cmaq2000}, provide an alternative source of pollutant concentration. Predictions from such models are provided either at the centroids of pixels or as an aggregate measure over the pixel on a regular square grid format, which generally span an extended spatial domain on a dense temporal scale (hourly or daily). However, the grid-cell or pixel sizes are often fairly large (typically 12 km x 12 km), providing crude spatial resolution. Additionally, these complex models do not use any observed measurements of the pollutant in the modeling process, and can often have significant bias associated with them. To account for these potential biases, various spatiotemporal modeling techniques have been developed that seek to calibrate output from such deterministic models using observed data on the pollutants. 
 
Most spatiotemporal calibration methods require temporal allignment between the observed data source and output from the deterministic model that needs to be calibrated, in addition to a somewhat dense spatial network of observed sites \citep{meiring1998, brown2001, li2008}. Additionally, these models do not address the issue of improving the spatial resolution of the large pixel sizes of deterministic model outputs, known as the ``change of support'' problem \citep{cressie1993}. While these methods work well when the pixel sizes are smaller, or if the process being modeled does not exhibit large variability over short distances, they are inadequate in modeling a process such an NO$_2$, which is known to vary considerably over short distances \citep{jerrett2004, who2003}. Other models do address this issue, but they are either purely spatial models \citep{fuentes2005}, or require fairly large number of spatial locations for accurately addressing the downscaling issue \citep{berrocal2010, alkuwari2013, chang2014}. 

\citet{gilani2016} developed a two-step modeling strategy, the Spatiotemporal Calibration and Resolution Refinement (SCARR) model, that allows calibrating estimates of a pollutant from a deterministic air quality model available in the form of grid-cell data, while also refining its spatial resolution, using two different sources of measured data that differ in their spatial and temporal resolutions. The modeling strategy was demonstrated by developing a space-time model using partial observations from three sources of data on the concentration of ambient \notwo over Connecticut in 1994, and its performance was tested using the remaining observations.  Additionally, for simplicity in the demonstrative example, the first step of the model was developed as a purely spatial model, without accounting for season as a predictor in the model.

In this paper, we extend the SCARR model and fit it to the same data sources using the complete set of observations to develop a space-time model to estimate the concentration of \notwo at a fine spatial and temporal resolution over the state of Connecticut for 1994 and 1995. Specifically, estimates of \notwo from the CMAQ model available in a grid-cell format with relatively large pixel sizes \mbox{(12 x 12 km)} are calibrated in space and time while also refining their spatial resolution using observations from two sources (\acid epidemiologic study data \citep{triche2002}, and US Environmental Protection Agency monitoring data \citepalias{epadata}) measured at different spatial and temporal resolutions. In this analysis, the SCARR model is extended in three ways: (a) the first step of the model is developed as a space-time model instead of a purely spatial model; (b) a parameter is included in the second step of the model that controls the influence of the estimated spatiotemporal calibration bias from the first step; and (c) additional covariates potentially correlated with atmospheric \notwo are included. The model is then used to predict the daily concentration of ambient \notwo for 1994 and 1995 over the entire state of Connecticut on a grid with a pixel size of 300 x 300 m. 

The remainder of the paper is organized as follows. Section 2 provides a description of the three different sources of available data on concentrations of NO$_2$, and the additional local covariates included in the model. Section 3 gives details on the two-step SCARR modeling strategy. Results for the fitted model are given in Section 4, while Section 5 provides predictions of \notwo for CT for 1994 and 1995 obtained from the fitted SCARR model. Finally, Section 6 provides some discussion and directions for future work.


\section{Data}\label{Paper2data}

\subsection{Sources of Data on \notwo Concentration} 
Data on the outdoor concentration of \notwo for the state of Connecticut (CT) for 1994 and 1995 are available from three different sources - predictions from the Community Multiscale Air Quality (CMAQ) model on a grid-cell format, and observed data from the \acid study and from EPA monitoring sites, both measured at different spatial and temporal resolutions.

\paragraph{\underline{CMAQ Model Data $(\ddot{Y}_1(\s,t))$}} \label{cmaqDesc}
Data from the Community Multiscale Air Quality (CMAQ) model version 4.7.1 were provided by the Atmospheric Sciences Research Center in Albany, New York \citepalias{byun2006, hogrefe2009, cmaq2000}. The model uses data from a meteorological forecast model, source emission inventories and chemistry transport modeling to predict hourly \notwo concentration on a regular grid over CT with each pixel of size 12 km x 12 km. These data have an extensive spatial coverage (over the entire state of CT) and are temporally dense, but provide estimates at the centroids of pixels with rather large sizes. Additionally, these estimates have not been calibrated to actual observed measurements of NO$_2$, and have systematic errors associated with them.

\paragraph{\underline{\acid Study Data  $(Y\!_2(\s,[t_\s]))$}} \label{acidDesc}
In the \acid study, 138 families were recruited from mothers delivering babies at seven Connecticut hospitals between 1993 and 1996 \citep{triche2002}. Of these, 129 families had outdoor \notwo concentrations measured at their residences by passive sampling using  Palmes Tubes \citep{palmes1976}. At the enrollment home visit, the \notwo monitoring tube was placed in an inverted funnel-shaped metal weather protector and hung from a tree branch or outdoor clothes line at least 5 ft above the ground and as close to the home as possible. The monitor was left in place for 10-14 days, and the cumulative concentration during that period was recorded. Point locations for the residences were obtained by geocoding each address against ESRI's\textsuperscript{\textregistered} \mbox{StreetsUSA} database \citep{ESRIstreet}; geocoding was unsuccessful for five locations, while two samples were excluded due to equipment contamination. The final analysis utilized 122 (94.6 percent) samples, all of which were collected at various times between March - December, 1994. These data are spatially dense but temporally sparse as they provide measurements for each residence only once a year, aggregated over a 10-14 day period. The index $[t_\s]$ for $Y_2$ indicates that data were observed over different lengths of time and at different time points for different \acid locations, $\s$.

\paragraph{\underline{EPA Data $(Y_{3}(\s,t))$}} \label{epaDesc}
The US Environmental Protection Agency (EPA) monitors atmospheric \notwo levels over four locations in CT (Bridgeport, East Hartford, New Haven, and Tolland) and two locations in southern Massachusetts (MA) (Chicopee and Springfield) on an hourly basis \citepalias{epadata}, which provide a rather accurate estimate of ambient \notwo at these locations. While these data are  temporally dense, they are spatially very sparse, with only six locations in \mbox{Connecticut} and southern Massachusetts. For each of the six EPA sites, the 24-hour mean daily \notwo concentration was calculated for the two years. 

\begin{figure}[h!t]
	\centering
	\begin{subfigure}[h!t]{0.7\textwidth}
		\includegraphics[width=\textwidth]{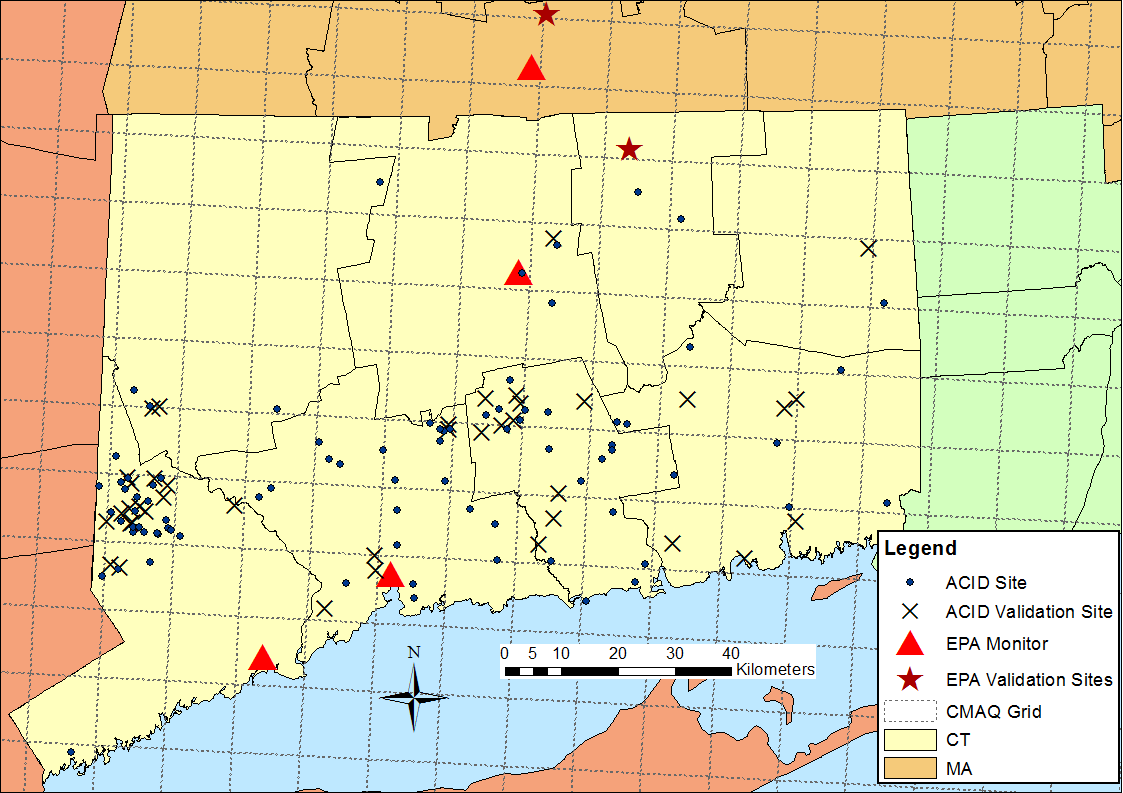}
		\caption{EPA and \acid sites with CMAQ grid overlaid.}
		\label{fig:SiteMap}
	\end{subfigure}
	\begin{subfigure}[h!t]{\textwidth}
		\includegraphics[width=\textwidth]{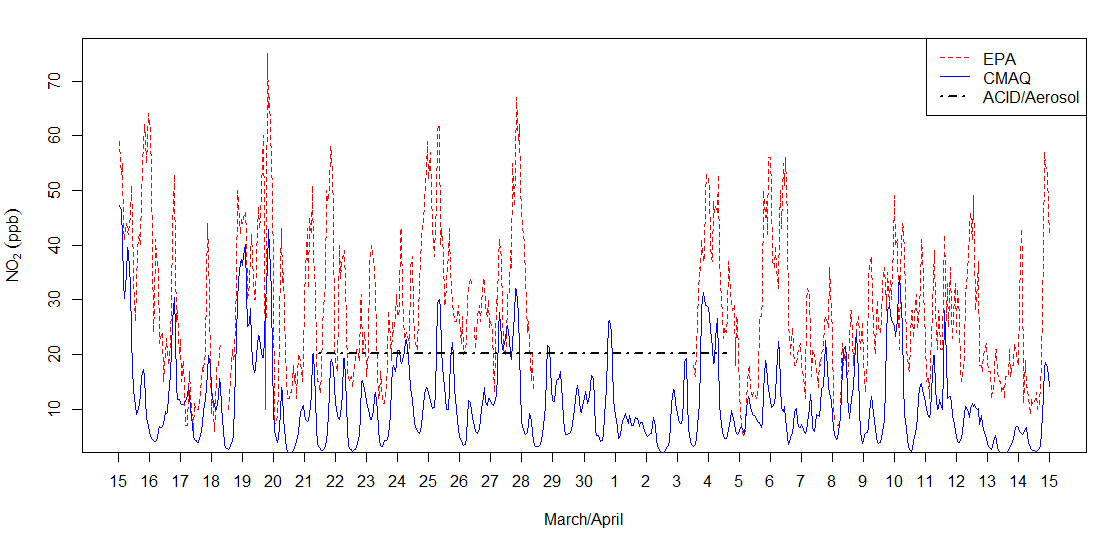}
		\caption[\notwo levels  for EPA site, CMAQ cell, and \acid site near New Haven, CT, March/April 1994.]{\notwo levels  for EPA site (red), CMAQ cell (blue), and \acid site (black) near New Haven, CT, March/April 1994. EPA data are missing for some time points, while \acid data provide a 13 day average measurement.}
		\label{fig:EHEPACMAQJ1994}
	\end{subfigure}
	\caption{Spatial and temporal distribution of study data.}
\end{figure}

The spatial distribution of the \acid and EPA sites with the CMAQ grid overlaid is given in Figure \ref{fig:SiteMap}, while Figure \ref{fig:EHEPACMAQJ1994} shows the variation in \notwo concentration over a 30 day period for the EPA monitor in New Haven, CT along with the CMAQ estimate and observations from a nearby \acid site for March/April 1994  (reproduced from \citet{gilani2016}). 

\subsection{Variables Used for Model Fitting} \label{Paper2Variables}
	
\paragraph{\underline{CMAQ Predictions at \acid and EPA Sites - $(\ddot{Y}\!_1(\s,t))$}} \label{Paper2CmaqAtAcid}

For each Acid/ Aerosol and EPA site, the closest CMAQ pixel centroid was identified, and a 24-hour  mean CMAQ \notwo concentration was calculated for the exact days that \notwo was measured at the \acid sites, and for 730 days in 1994 and 1995 at the EPA sites. 

\paragraph{\underline{Traffic Measurements}} \label{Paper2traffic}

The Department of Transportation for Connecticut, Massachusetts and New York record annual traffic volume in the form of average daily traffic (ADT) for interstates and numbered highways \citep{ctdot2000, madot, nydot}. Following the approach outlined by  \citet{holford2010} and  \citet{skene2010} and using traffic data for 1994 and 1995 provided by the Connecticut, Massachusetts and New York Departments of Transportation, we divided a line file of interstates and numbered highways  into approximately 50-meter segments, each of which had an associated ADT count. We defined the midpoint of each segment and calculated a measure of traffic volume (TV) on that segment as the product of segment length and ADT. The contribution of a segment to \notwo concentration at any location can be expressed as the product of TV and a dispersion function of distance and direction between the segment midpoint and the site. \citet{holford2010} found that the dispersion function can be effectively estimated by a step-function with steps of distance 0-0.5 km, 0.5-1 km, 1-2 km, 2-3 km, 3-4 km, 4-5 km and 5-6 km. Beyond 6 kms, the effect of vehicular traffic on \notwo was not statistically significant \citep{holford2010, skene2010}. 

To estimate the dispersion function, circular buffers of radii 500 m, 1 km, 2 km, 3 km, \mbox{4 km,} 5 km and 6 km were created around each \acid and EPA site, and the total traffic volume (TTV) within each concentric buffer ring was calculated by summing the contribution of all point sources within the buffer ring and dividing by 10,000. This gave a measure of 10,000 vehicle-kilometers per day for a given distance range - e.g., a total traffic volume value of 3 at an \acid site for a buffer ring of 0.5-1 km would indicate an average of 30,000 vehicle-kilometers traveled per day within that buffer ring.

\paragraph{\underline{Land Use Covariates}} \label{Paper2LU}

Land use data for Connecticut were obtained from the United States Geological Survey (USGS) `National Land Cover Dataset' (NLCD) for the year 1992 \citep{lu}. The data were stored as a raster file with 30-meter pixels and classified into 17 categories: open water, low intensity residential, high intensity residential, commercial/industrial, bare rock, quarries, transitional barren, deciduous forest, evergreen forest, mixed forest, shrubland, orchards, pastures, row crops, urban/recreational grasses, woody wetlands and emergent herbaceous wetlands.

For each \acid site, we used the ``Intersect Points with Raster" tool from the Geostatistical Modeling Environment \citep{gme, R} to count the number of pixels of each land use category within circular buffer rings of size 0-0.5 km, 0.5-1 km and 1-2 km. This total was then multiplied by 900 (area in m$^2$ of a 30 m pixel) and divided by 10,000 to give the area in hectares of each land use category (LUC) within the three buffer rings.

\paragraph{\underline{Population Density}} \label{Paper2PopDen}

The 1990 mid-year census tract population and polygon shape file for census tracts was obtained from the US Census Bureau \citep{censusShape, census1990}. Population density (per square mile) was assigned to each residence as the mid-year population of the census tract divided by the area of the census tract in square miles.

\paragraph{\underline{Elevation}} \label{Paper2Elevation}

Elevation above sea level, in meters, for each residence was extracted as the raster cell value from the USGS `National Map'  for the year 2005 \citep{gesch2007, gesch2002}.

\paragraph{\underline{Season}} \label{Paper2Season}

Concentrations of \notwo follow a seasonal cycle, and are generally higher in the winter months as compared to the summer. To capture the effect of season, a trigonometric and linear function of date was included in the model. Day of the Year Ratio (DYR) was defined as the midpoint of the days that \notwo was monitored at an \acid residence divided by 365, and four covariates were calculated: sin($2.\pi.$DYR), cos($2.\pi.$DYR), sin($4.\pi.$DYR), cos($4.\pi.$DYR). 

%


\section{Model} \label{Paper2ModelingStrategy}

As outlined by \citet{gilani2016}, the spatiotemporal calibration and resolution refinement (SCARR) model to calibrate the predictions of the concentration of ambient \notwo from the CMAQ model and to improve its spatial resolution is developed in two steps. The first step, Calibration and Spatial Refinement, uses the spatially dense but temporally sparse \acid data along with other publicly available data on various local covariates related to the modification and dispersion of NO$_2$ to calibrate the CMAQ predictions over space and time, while also improving their spatial resolution. This step provides a continuous space representation of the artificially discrete CMAQ data, and estimates the additive and multiplicative calibration bias of the CMAQ data. The second step, Spatiotemporal Calibration using Dynamic Space-Time modeling, improves the temporal resolution of Step I by using the temporally dense but spatially sparse EPA data for estimating the temporal evolution of the additive and multiplicative calibration constants at a finer temporal resolution (daily). The SCARR model assumes that there is large scale spatial variation in the concentration of the pollutant over the region $\mathcal{S}$ of interest as well as small scale variation. The CMAQ data capture the large scale variation but not the small scale variation. It also assumes that the small scale spatial variation in pollutant concentration is due to local factors such as land use type, traffic density, population density, elevation,  and that this small scale spatial variability is similar between the shorter time interval (1 day) of the EPA data and the longer time durations (10-14 days) of \mbox{the \acid data.} 

	
\subsection{Step I} \label{Paper2StepIModel}

We fitted a model to calibrate and refine the granularity of the daily mean CMAQ \notwo concentration using \acid \notwo data. A spatiotemporal calibration and refinement model, as described by \citet{gilani2016}, is specified by:
\begin{equation}\label{eqn:Paper2IEM}
\overline{Y}\!_2(\s,t) = \frac{\bigintsss_{[t_\s]} \big(\mathbf{X}'(\s,t)\boldsymbol{\beta}  + \mathbf{G}'(\s,t)\boldsymbol{\lambda} +  \ddot{Y}\!_1(\s,t)\gamma  \big)~dt}{\|[t_\s]\|} + \nu(\s,t) + \epsilon (\s,t),
\end{equation}
where $\overline{Y}\!_2(\s,t)$ is the concentration of \notwo from the \acid data observed at location $\s$ averaged over the 10-14 day time interval $[t_\s]$; $\mathbf{X}$ are the covariates not related to dispersion, and includes a column of ones for the intercept; $\mathbf{G}$ is the integral of the product of the dispersion related covariates and the intensity $\mathbf{Z}$ at the source  located within a local neighborhood; $\ddot{Y}\!_1(\s,t)$ is the estimate from the CMAQ data at the pixel centroid nearest to location $\s$; and $\epsilon (\s,t)$ are independent mean zero Gaussian random errors. Residual spatiotemporal dependence between the sites is modeled using the hierarchical error term, $\nu(\s,t)$, where \mbox{$\boldsymbol{\nu} \sim$ MVN$(\mathbf{0}, \sigma^2\mathbf{\Sigma})$}. The covariates $\mathbf{G}$ related to dispersion include total traffic volume (TTV) and land use categories (LUC), where the dispersion is modeled using a step function, which is estimated through the model. For the covariates $\mathbf{X}$ not related to dispersion, \citet{gilani2016} included only one variable, population density, in their model. In our analysis, we further include elevation, as well as a trigonometric function of time to capture the  seasonal variation in \notwo concentrations. Given the temporal sparsity of the \acid data, it was difficult to accurately estimate a space-time covariance matrix $\boldsymbol{\Sigma}$. Additionally, aggregation over the 10-14 day time interval accounted for the short-range temporal dependence at a given site $\s$. We therefore assumed that most of the residual dependence between sites was spatial in nature.  We considered both spatially dependent and independent error models. For the spatially dependent error models, spherical, exponential and  Mat\'{e}rn covariance functions were used to model $\mathbf{\Sigma}$, whereas for the spatially independent error model, the hierarchical error term was removed and a linear regression framework was utilized.

The initial model included all the variables, and the traffic covariates were selected by a backward approach, giving preference to buffer rings closer to the residences. For example, to include the TTV for the buffer 1.0-2.0 km, the buffers 0-0.5 km and 0.5-1.0 km must also be included in the model. The same strategy was adopted for land use categories. Nested traffic and land use buffer models were compared using $F$-tests, and buffer variables not contributing significantly to the model (at the 5\% level) were removed. Leave-one-out cross-validation was also performed on the best few models, and preference was given to models for which the prediction sum of squares was closer to the error sum of squares. 

\subsection{Step II} \label{Paper2StepIIModel}

We fitted a dynamic space-time model for temporal calibration of the spatially refined CMAQ estimates from Step I. Assumption 3 of the modeling strategy allows us to generalize the results of Step I to calibrate estimates of the CMAQ data on a finer temporal resolution in Step II using the temporally dense EPA data. Let $\s_1, \ldots, \s_6$ represent the spatial location of the six EPA monitor sites. The spatiotemporal additive calibration bias $\widetilde{C}(\s_i,t)$ at location $\s_i, i = 1, \ldots 6,$ from Step I is defined as \mbox{$\widetilde{C}(\s_i,t) = \mathbf{X}'(\s_i,t)\boldsymbol{\hat{\beta}}  + \mathbf{G}'(\s_i,t)\boldsymbol{\hat{\lambda}}$}, where   $\boldsymbol{\hat{\beta}}$ and $\boldsymbol{\hat{\lambda}}$ are vectors of parameters estimated in the first step. Note that the tilde (~$\widetilde{}$~) on $C$ signifies that this variable was calculated using parameters estimated in Step I of the model. We further define \mbox{$\mathbf{Y}\!_3(t) = \big(Y_3(\s_1,t),  \ldots, Y_3(\s_6,t)\big)'$}; ~ $\mathbf{\ddot{Y}}\!_1(t)$ a diagonal matrix with diag$\big(\mathbf{\ddot{Y}}\!_1(t)\big)= \big(\ddot{Y}\!_1(\s_1,t), \ldots, \ddot{Y}\!_1(\s_6,t)\big)$; and 
$\mathbf{\widetilde{C}}(t) = \big(\widetilde{C}(\s_1,t),  \ldots, \widetilde{C}(\s_6,t)\big)'$. The \emph{observation equation} for  a dynamic spatiotemporal model describing the relationship between $\mathbf{Y}\!_3(t)$ and $\mathbf{\ddot{Y}}\!_1(t)$  in \citet{gilani2016} is extended by adding a parameter $\beta_c$, which controls the influence of $\mathbf{\widetilde{C}}(t)$ in the temporal calibration of Step II. The modified observation equation is thus given by 
\begin{equation} \label{eqn:Paper2obs}
\mathbf{Y}\!_3(t) = \mathbf{A}(t) +  \beta_c\mathbf{\widetilde{C}}(t) + \hat{\gamma}\mathbf{\ddot{Y}}\!_1(t) + \mathbf{Z}(t), \hspace{1in} t = 1, 2, \ldots, 730, 
\end{equation}
where $\mathbf{Z}(t) = \big(Z(\s_1,t), \ldots, Z(\s_6,t)\big)'$ is a multivariate white noise time series with the $Z(\s_i,t),$ $i=1,\ldots,6$ mutually independent $N\big(0, \sigma^2_{Z}(\s)\big)$ variates. $\mathbf{A}(t) = \big(A(\s_1,t), \ldots, A(\s_6,t)\big)'$ is a stochastic process whose evolution over time is described by the \textit{state equation}
\begin{equation}\label{eqn:Paper2state1}
\mathbf{A}(t) - \mu_{A} = \boldsymbol{\Psi}_{A}(\mathbf{A}(t-1)- \mu_{A}) + \mathbf{F}\boldsymbol{\xi}(t), 
\end{equation} 
where $\boldsymbol{\xi}(t) = \big(\xi(\s_1,t),  \ldots, \xi(\s_6,t)\big)'$ is a multivariate stochastic process with $\xi(\s_i,t), i=1,\ldots,6,$ mean zero Gaussian variates with variance $\sigma^2_{A}(\s)$.  $\boldsymbol{\Psi}_A$ is a diagonal matrix with $diag(\boldsymbol{\Psi}_A) = \big(\psi_A(\s_1), \ldots, \psi_A(\s_6)\big)$, where $\psi_A(\s_i), i=1,\ldots,6$ are autoregressive parameters lying in the range 0~--~1. $\beta_c$ is a parameter that is constant in space and time.

In the model described above, $\mathbf{A}(t) +  \beta_c\mathbf{\widetilde{C}}(t)$ provides the additive calibration bias on the finer temporal scale, while $\hat{\gamma}$, estimated in Step I, provides the multiplicative calibration bias. The matrix $\mathbf{F}$ in equation \ref{eqn:Paper2state1} can be used to model the spatial correlation in the additive bias to arrive at an integrated space-time model. However, for this example, with only six locations spread over relatively large distances, accurately estimating the spatial correlation structure of $\mathbf{A}(t)$ is not possible. We therefore assumed spatial independence for $A(\s_i,t)$ between the six sites. Additionally, we assumed that the six sites have identical parameters  $\sigma^2_Z,~\sigma^2_A,~ \psi_A,$ and $\mu_A$, allowing us to pool across the six sites to estimate $A(t)$ common to the entire study region. In this setting,  $\beta_c\, \widetilde{C}(\s,t)$ captures the spatial and two-week temporal variability in the additive calibration bias, while $A(t)$ provides the finer-scale temporal evolution of the additive calibration bias that is common for all sites. Under these assumptions, equations \ref{eqn:Paper2obs} and \ref{eqn:Paper2state1} can be rewritten in vector notation as
\begin{eqnarray} \label{eqn:Paper2obsExVec}
\mathbf{Y}_{3}(t)& = &A(t) \mathbb{1}_6+ \beta_c\mathbf{\widetilde{C}}(t) + \hat{\gamma}\mathbf{\ddot{Y}}\!_{1}(t) + \sigma_zZ(t)\mathbf{I}_6 \\  \label{eqn:Paper2stateExVec}
A(t)  - \mu_A &=& \psi_A (A(t-1) - \mu_A) + \sigma_A\xi(t) \hspace{0.5in} t = 1, 2, \ldots, 730,
\end{eqnarray}
where $Z(t)$ and $\xi(t)$ are standard Gaussian variates, $\mathbb{1}$ a vector of 1's and $\mathbf{I}$  the identity matrix. 

\indent The dynamic space-time model outlined in equations \ref{eqn:Paper2obsExVec} and \ref{eqn:Paper2stateExVec} was fitted to data using the Kalman filter \citep{kalman1960} \citep[see][for details]{gilani2016}. Both maximum likelihood and Bayesian techniques using Markov Chain Monte Carlo simulations were utilized to estimate the parameters $\sigma^2_{Z}$, $\sigma^2_{A}$, $\psi_{A}$, $\mu_{A}$ and $\beta_c$ using the package \pkg{dlm} \citep{petris2010} in \proglang{R} statistical package. Estimates produced by the two methods for the full model were quite similar, and due to the faster computational speed of maximum likelihood estimation (MLE), further model fitting was conducted using MLE. 

To test the fit of the two-step SCARR model, we calculated the coefficient of correlation and the empirical mean squared error (MSE) for each of the six sites, comparing the EPA observations with predictions from the SCARR and CMAQ models. We also fit the final SCARR model at the 122 \acid locations (details on fitting the full model at a new location $\s'$ are given in Section 5) and calculated the mean square prediction error (MSPE) for the SCARR model predictions and the CMAQ model estimates at these sites. Diagnostic plots were used to check for violations of model assumptions.


\section{Results} \label{Paper2Results}

	
\subsection{Step I} \label{Paper2StepIResults}

\begin{figure}[h!t]
	\centering
	\begin{subfigure}[h!t]{0.49\textwidth}
		\includegraphics[width=\textwidth]{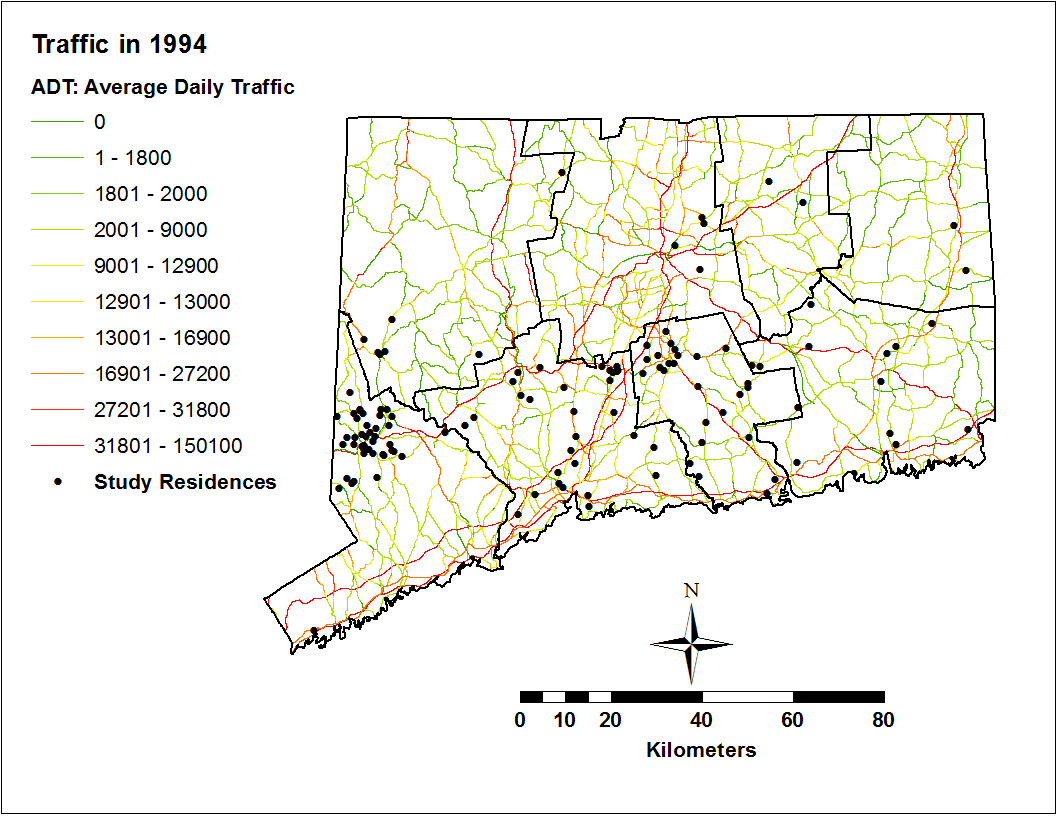}
		\caption{\footnotesize{Average daily traffic (ADT) counts on interstate and numbered highways in CT, 1994 \citep{ctdot2000}.}}
		\label{fig:Paper2adt}
	\end{subfigure} 
	\begin{subfigure}[h!t]{0.49\textwidth}
		\includegraphics[width=\textwidth]{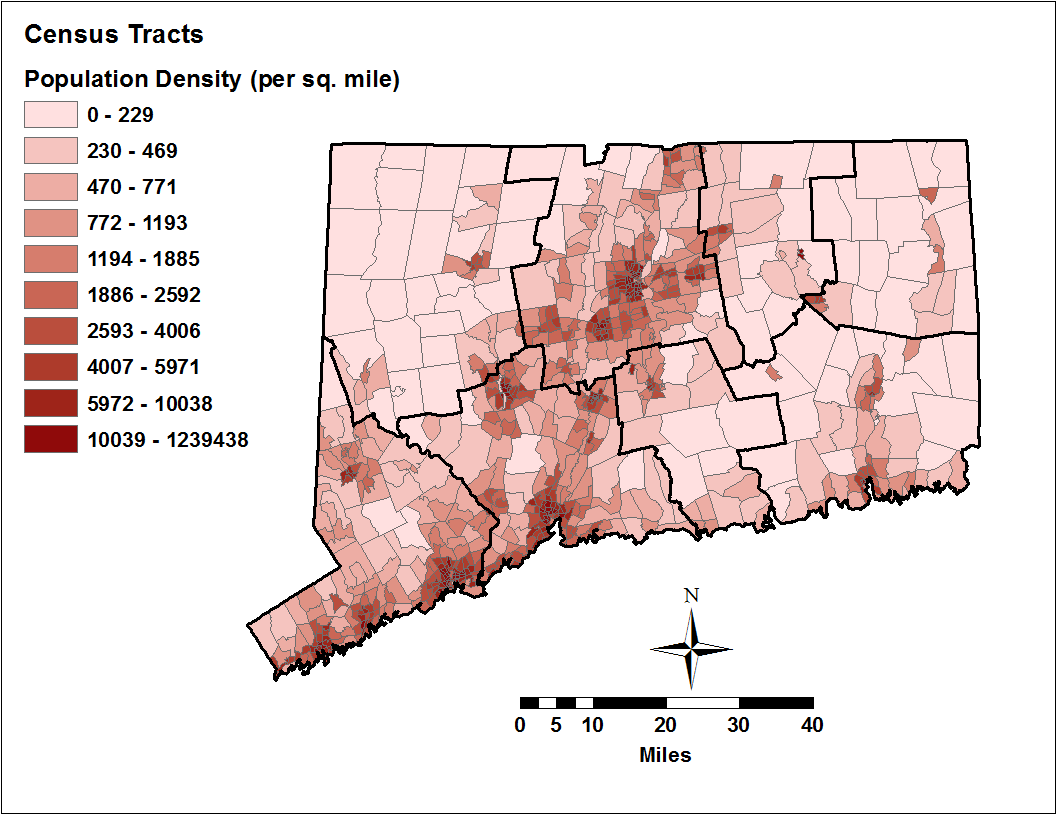}
		\caption{\footnotesize{Population Density (per square mile), CT census tracts, 1990 \citep{censusShape, census1990}.}\\}
		\label{fig:Paper2popden}
	\end{subfigure} 
 \par \bigskip
	\begin{subfigure}[h!t]{0.49\textwidth}
		\includegraphics[width=\textwidth]{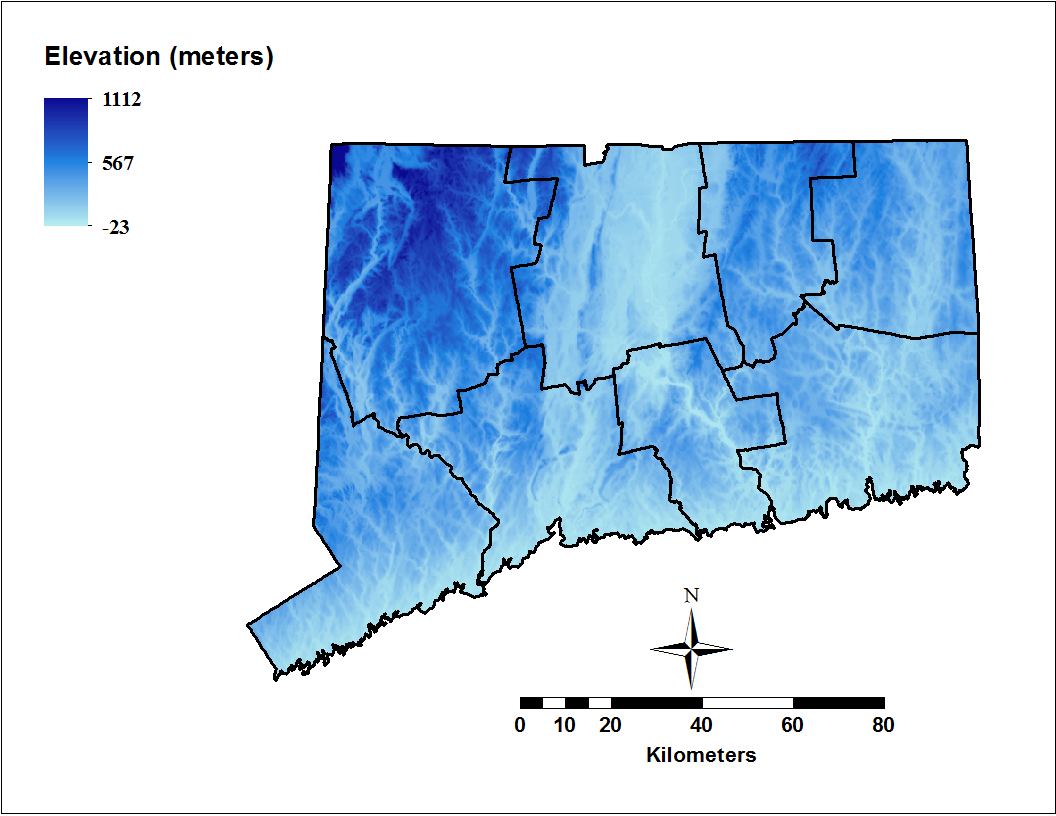}
		\caption{\footnotesize{Elevation (meters) in CT  \citep{gesch2007, gesch2002}.}}
		\label{fig:Paper2elevation}
	\end{subfigure}
	\begin{subfigure}[h!t]{0.49\textwidth}
		\includegraphics[width=\textwidth]{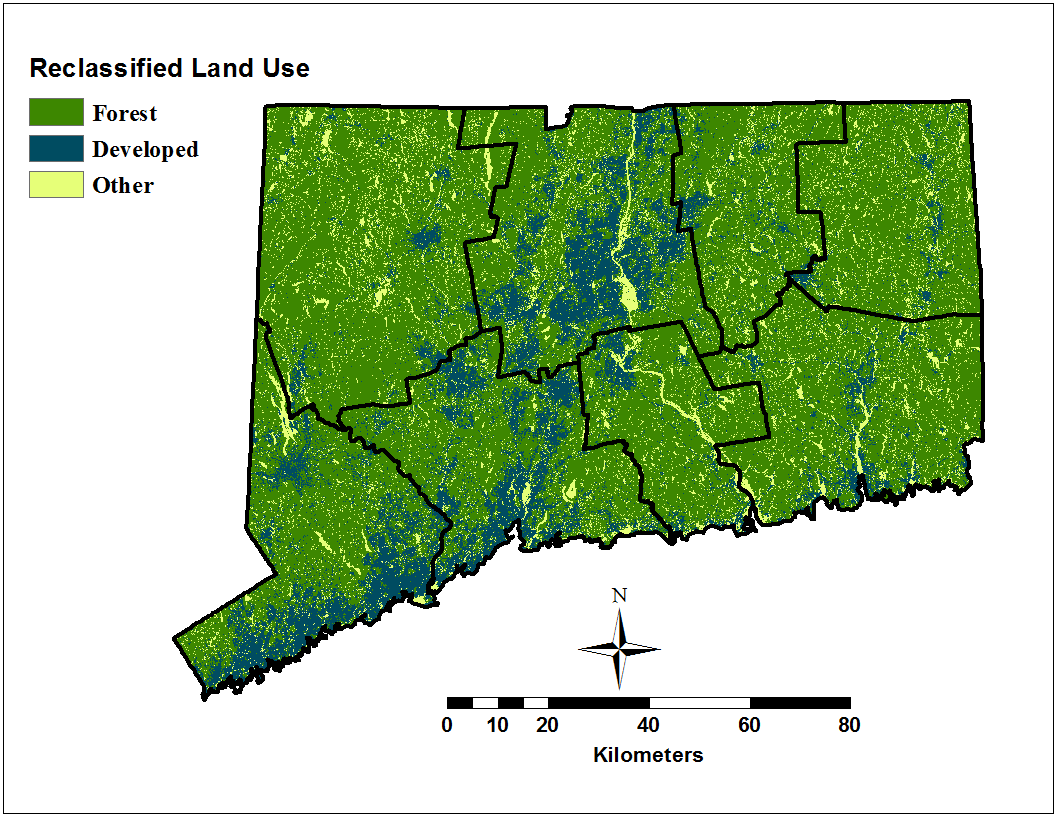}
		\caption{\footnotesize{Reclassified land use in CT, 1992 \citep{lu}.}}
		\label{fig:Paper2lu}
	\end{subfigure} 	
	\caption{Geographical distribution of spatial covariates.} \label{Paper2Maps}
\end{figure}

Table \ref{tab:Paper2Sum1} gives a summary of all covariates used to develop the spatial calibration and refinement model.  Figures \ref{fig:Paper2popden} and \ref{fig:Paper2elevation} show the geographic distribution of population density in Connecticut for 1990 and elevation in 1992, respectively, while Figure \ref{fig:Paper2adt} shows the distribution of ADT for interstates and numbered highways for 1994. 

\begin{table}[!htp]
\caption{Summary statistics for Step I model covariates.}\label{tab:Paper2Sum1}
\small
\begin{center}
 \begin{tabular}{l c c}
\toprule \toprule

   \bf Variable                         &      \bf Mean (SD)$^*$  &       \bf Range \\
      \midrule

\acid \notwo (ppb)$^a$    & 13.6    (5.28) &  4.39  -  33.1 \\
CMAQ \notwo (ppb)$^b$  & 10.6    (4.75) &  2.51  -  24.0 \\
Traffic Density (traffic volume/km$^2$) \\ 
~~~\emph{Buffer ring (km)}$^c$ \\
~~~~~ 0.0 - 0.5            & 0.89 (1.92)  & 0.00 - 12.5    \\
~~~~~ 0.5 - 1.0            & 1.00 (1.52)  & 0.00 - 8.00    \\
~~~~~ 1.0 - 2.0            & 1.14 (1.31)  & 0.00 - 5.95   \\
~~~~~ 2.0 - 3.0            & 0.99 (1.02)  & 0.02 - 4.40   \\
~~~~~ 3.0 - 4.0            & 1.11 (1.08)  & 0.05 - 5.38   \\
~~~~~ 4.0 - 5.0            & 0.91 (0.71)  & 0.06 - 3.12   \\
~~~~~ 5.0 - 6.0            & 0.89 (0.62)  & 0.05 - 2.50   \\
Land Use Density (hectares/km$^2$)   \\
~~~\emph{Category/Buffer ring (km)}$^d$\\
~~~Developed$^e$ \\
~~~~~ 0.0 - 0.5            &19.89  (15.82) & 0.17 - 49.27  \\
~~~~~ 0.5 - 1.0            & 53.07  (42.34) &  0.40 - 146.7  \\
~~~~~ 1.0 - 2.0            & 186.65 (141.35) &  2.64 - 558.3 \\
~~~Forest$^f$ \\
~~~~~ 0.0 - 0.5           & 8.39   (5.01) & 0.08 - 15.91 \\
~~~~~ 0.5 - 1.0           & 26.78  (13.88) & 0.55 - 48.89 \\
~~~~~ 1.0 - 2.0           & 113.99  (47.69) & 8.54 - 193.9 \\
~~~Other$^g$ \\
~~~~~ 0.0 - 0.5           & 0.41   (0.47) & 0.00 -  3.07~ \\
~~~~~ 0.5 - 1.0           & 1.38   (1.39) & 0.12 -  10.47 \\
~~~~~ 1.0 - 2.0          & 5.95   (5.67) & 0.50 -  42.84 \\
Population Density (population/mi$^2$)  &
                                     1682 (2284)    & 86.3 - 11529 \\
Elevation (m)               & 106   (68.5)    & 1.60 -  331.57 \\
 \bottomrule
\end{tabular} \end{center}
\caption*{\footnotesize{
\!$^{*}N = 122$ for all covariates. \\
$^a$ 24 hr mean \notwo concentration sampled outside 122 homes over a 10-14 day period.\\
$^b$ 24 hr mean \notwo concentration estimated from the CMAQ model for the same 10-14 day period as the \acid residence.\\
$^c$ Total daily traffic volume (in 10,000 vehicle-km) divided by area (in km$^2$) of buffer ring. \\
$^d$ Total hectares of land use category  divided by area (in km$^2$) of buffer ring.\\
$^e$ Developed includes low intensity residential, high intensity residential, and commercial/industrial.\\
$^f$ Forest includes deciduous, evergreen and mixed forests, pastures, row crops, and urban/recreational grasses. \\ 
$^g$ Other includes open water, bare rock, quarries, transitional barren, shrubland, orchards, woody wetlands and emergent herbaceous wetlands.\\}}
\end{table}

\begin{figure}[h!t]
	\centering
	\begin{subfigure}[h!t]{0.7\textwidth}
		\includegraphics[width=\textwidth]{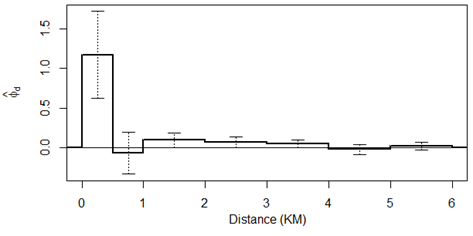}
		\caption{}
		\label{fig:Paper2dispIso}
	\end{subfigure}
	\par \bigskip
	\begin{subfigure}[h!t]{0.7\textwidth}
		\includegraphics[width=\textwidth]{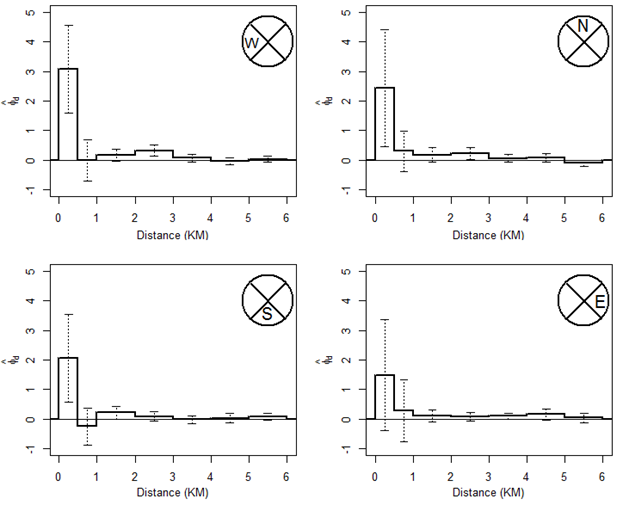}
		\caption{}
		\label{fig:Paper2dispQuad}
	\end{subfigure}
	\caption{Estimated (a) isotropic and (b) anisotropic step dispersion functions for total traffic volume (TTV) based on distance $d$ and direction from a residence.}
\end{figure}

We explored the shape of the dispersion function relating \notwo to total traffic volume by fitting a model using just the TTV buffer covariates. The estimated dispersion function is given in  Figure \ref{fig:Paper2dispIso}. To check for small-scale spatial anisotropy in the dispersion of \notwo related to traffic, we divided the traffic buffers into four quadrants and calculated the total traffic volume within each quadrant for each buffer ring. Figure \ref{fig:Paper2dispQuad} shows the step dispersion function in each direction, estimated by fitting the directional traffic buffer covariates in separate models for each direction. The shape of the four directional dispersion functions appear very similar to each other and to the estimated isotropic dispersion function in \mbox{Figure \ref{fig:Paper2dispIso},} suggesting small scale spatial isotropy in the dispersion of NO$_2$. We therefore used the omnidirectional total traffic volume covariates in developing the model.  

Based on previous results \citep{skene2010} and exploratory analysis, we reclassified the land use categories  into developed (including low intensity residential, high intensity residential and commercial/industrial), 
forest (including deciduous, evergreen and mixed forests, pastures, row crops, and urban/recreational grasses), and other (all other categories). Figure \ref{fig:Paper2lu} shows the spatial distribution of the reclassified land use categories for Connecticut in 1992.  

An initial analysis of the land use categories displayed a high degree of negative correlation between the ``developed'' and ``forest'' categories (e.g. -0.92 between ``developed \mbox{0.5-1 km}'' and ``forest 0.5-1 km'' rings). Additionally, within each of the three categories, we observed very high positive correlations between the buffer rings (e.g. 0.91 between 0-0.5 km and \mbox{0.5-1 km} ``forest'' rings). This raised the concern of multicollinearity if all the land use category covariates were included in the model together. In fact, inclusion of all of these covariates in the model resulted in unstable and unusual parameter estimates. Due to the high degree of correlation between the different buffer rings within the categories, the buffer rings for each category were combined to derive a single covariate for each category from 0-2 km. The new categories ``developed 0-2 km'' and ``forest 0-2 km'' had a correlation coefficient of -0.90, and including both categories in the model did not provide significant improvement in the model as compared to including just one covariate. Including these two covariates separately in the model, while keeping all other covariates fixed, provided almost exactly the same parameter estimates, with the sign reversed. Therefore, in the final model \mbox{``forest 0-2 km'}' alone was included. The \mbox{``other 0-2 km''} land use category did not significantly improve the model in the presence of \mbox{``forest 0-2 km'',} and was therefore not included in the final model. 

We fit a spatially independent error model using a linear regression framework to identify the best subset of variables to include in the model. Omnidirectional and four-directional  variograms of the residuals (not shown) revealed no evidence of spatial correlation in the residuals, which justified the assumption of spatial independence of the errors. However, we fitted spatially dependent error models as well. The covariates from the final regression model were included in a spatially dependent error model with the spherical, exponential, and Mat\'{e}rn covariance functions to model the spatial dependence in the errors using \pkg{PROC MIXED} in \proglang{SAS 9.3} (SAS Institute, Cary, NC). These models provided similar parameter estimates as the linear regression model, arriving at the maximum likelihood when the estimated covariance parameters provided an effective range of zero.

\begin{figure}[h!t]
	\centering
		\includegraphics{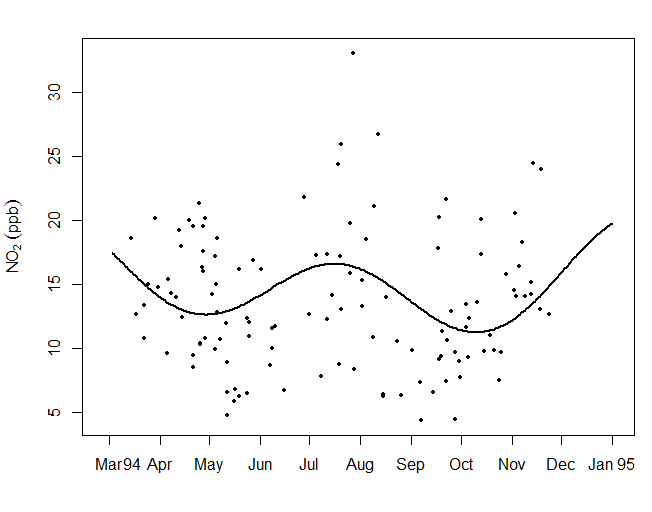}
	\caption{Observed concentrations of \notwo along with the estimated trigonometric and linear function of date to represent season.}
	\label{fig:Paper2Season}
\end{figure}

In the presence of the other covariates, elevation did not appear to improve the fit of the model. CMAQ NO$_2$, population density and the trigonometric and linear function of date were statistically significant, and were included in the final model. None of the TTV buffer rings further than 2 km of the \acid sites were statistically significant, and were removed from the model. A summary of the covariates in the final model and their parameter estimates is given in \mbox{Table \ref{tab:Paper2parEst}}, while Figure \ref{fig:Paper2Season} shows the observed concentrations of \notwo from the \acid data, along with the estimated trigonometric and linear function of date. 

%

\begin{table}[!ht]
\caption{Variables in the final model for Step I, along with their parameter estimates and 95\% confidence intervals. }\label{tab:Paper2parEst}
\footnotesize
\begin{center}
 \begin{tabular}{l c c}
\toprule \toprule

   \bf Parameter                         &      \bf Estimate  &       \bf (95\% CI) \\
      \midrule
$\boldsymbol{\beta}$:	\\
    ~~Intercept                                       & 11.891  	& (8.894, 14.888)  \\
    ~~Population Density (10,000)        & 5.508             	& (2.656, 8.359)\\
    ~~Season\\
    ~~~~~sin($2.\pi.$DYR)                      & 1.245  	& (0.424,  2.065)\\
    ~~~~~cos($2.\pi.$DYR)                     & 1.727 		& (-0.207,  3.662)\\
    ~~~~~sin($4.\pi.$DYR)                      & 1.792  	& (0.883,  2.702)\\
    ~~~~~cos($4.\pi.$DYR)                     & 2.706  	& (1.685,  3.728) \\
$\boldsymbol{\lambda}$:	\\
    ~~Total Traffic Vol. (10,000 v-km)   \\
    ~~~~~~~ 0.0 - 0.5 km                           & 0.851  	& (0.496,  1.207) \\
    ~~~~~~~ 0.5 - 1.0 km                           & -0.138 	& (-0.310,  0.035) \\
    ~~~~~~~ 1.0 - 2.0 km                           & 0.084  		& (0.031,  0.136)\\
  
    ~~Land Use (1,000 hectares)        &                         &   \\
     ~~~~~\emph{Forest} \\
     ~~~~~~~ 0.0 - 2.0 km                    & -5.430               & (-7.606, -3.254)\\

   $\gamma$:   \\
     ~~CMAQ \notwo                                & ~0.487 	& (0.333,  0.642)\\
    
     	\\
     $R^2$			              & 0.777                     & \\
     $R^2$ (Adjusted)                       & 0.757                  & \\
     RMSE				    & 2.60 \\
     RMSPE$^*$                                 & 2.77\\
 \bottomrule
\end{tabular} \end{center}
\caption*{\footnotesize{
\!$^*$ Leave-one-out cross-validation root mean squared prediction error.\\
Final model: Acid/Aerosol \notwo =  Population Density + Season + TTV (0-0.5 km, 0.5-1.0 km, 1.0-2.0 km) + LUC (Forest: 0-2 km) + CMAQ \notwo} }
\end{table}

\setkeys{Gin}{width=.7\textwidth}


\subsection{Step II} \label{Paper2StepIIResults}

\begin{table}[!]
\caption{Summary statistics for \notwo concentration (ppb) for EPA and CMAQ data at six EPA sites, along  with the spatiotemporal additive bias $\widetilde{C}$ at these sites, calculated using parameter estimates from Step I.} \label{tab:Paper2summaryEPA}
\small
\centering
\begin{tabular}{l ccc cc c c}
\toprule \toprule

 & \multicolumn{2}{c}{\bf{EPA} ($Y_3$)}  &  \multicolumn{2}{c}{\bf{CMAQ} ($\ddot{Y}_1$)} & & $\mathbf{\widetilde{C}}$\\
\cmidrule(lr){2-3} \cmidrule(lr){4-5} \cmidrule(lr){7-7} 
\bf Site &   \bf Mean (SD)  & \bf  Range  &  \bf Mean (SD)  & \bf  Range & $r^a$ &  \bf Mean \\
\midrule
Bridgeport  &  25.3 (10.5) & 5.3 - 74.2 & 20.2 (9.7) & 3.9 - 55.7 & 0.693 & 16.7 \\
Chicopee    &  15.9 ~(9.8)   & 1.0 - 71.7 & ~9.1 (7.6)  & 1.0 - 50.1 & 0.749 & 10.8 \\
E. Hartford &  18.5 ~(9.8)   & 1.0 - 63.4 & 16.1 (8.6) & 2.3 - 53.3 & 0.763 & 13.6 \\
New Haven & 27.5 (10.2) & 5.9 - 74.9 & 19.7 (9.0) & 3.5 - 55.0 & 0.658 & 24.6 \\
Springfield  & 26.0 (11.1) & 4.1 - 84.0 & 14.6 (9.0) & 2.0 - 9.90 & 0.681 & 20.2 \\
Tolland        & ~9.5 ~(6.5)   & 1.0 - 41.9 &  ~9.1 (7.0) & 1.1 - 46.6 & 0.819 & 6.13 \\

\bottomrule 
\end{tabular}
\caption*{
$^a$ Coefficient of correlation between observed EPA data and CMAQ data. }
\end{table}

Table \ref{tab:Paper2summaryEPA} provides summary statistics for \notwo concentrations at the six sites for the EPA and CMAQ data, along with the correlation between these two sources of data at each location. It also gives the estimate of the spatiotemporal additive bias $\widetilde{C}(\s,t)$ at each site, calculated using parameter estimates from Step I. The EPA site at Springfield had observations for all 730 days, while Tolland had observations only for 369 days between October 1994 and November 1995. The other four sites had EPA data missing for a few days, with the available data ranging between 697 - 727 days. CMAQ data generally underestimates \notwo concentrations, with the difference most pronounced for New Haven and Springfield. The additive bias $\widetilde{C}(\s,t)$ from Step I captures this difference, providing highest estimates for these two sites.  

\begin{figure}[!ht]
	\centering
		\includegraphics{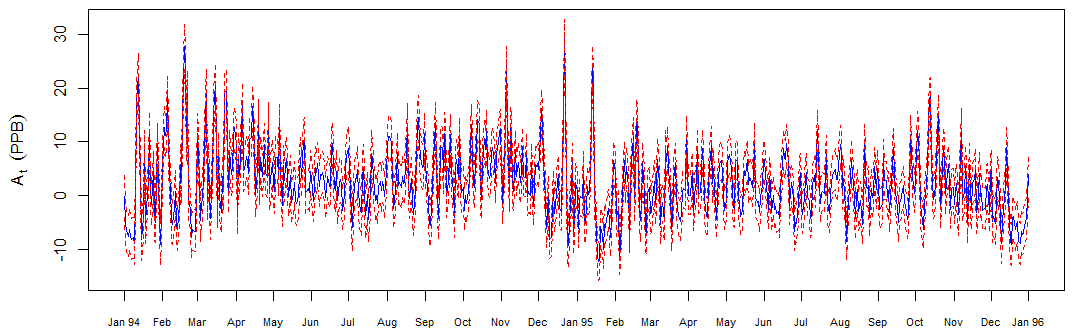}
	\caption[Smoothed estimate of $A(t)$ with 95\% confidence interval.]{Smoothed estimate of $A(t)$ (solid blue) with 95\% CI (dotted red).}
	\label{fig:Paper2at}
\end{figure}

\begin{sidewaysfigure}[!]
	\centering
	\begin{subfigure}{\textwidth}
		\includegraphics{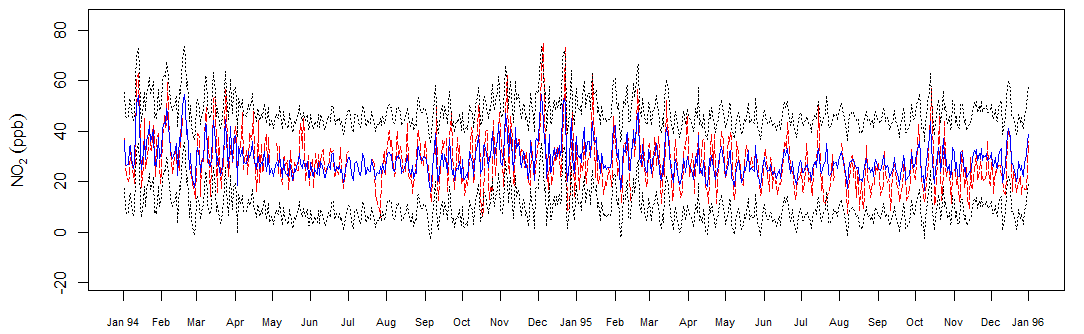}
	\caption[Observed EPA and SCARR fitted \notwo conc. at New Haven, CT, with 95\% confidence intervals.]{Observed EPA (red) and SCARR fitted (blue) \notwo conc. at New Haven, CT, with 95\% confidence intervals (black dotted).}
	\label{fig:Paper2fittedNewHavenCI}
	\end{subfigure}
	\begin{subfigure}{\textwidth}
	\includegraphics{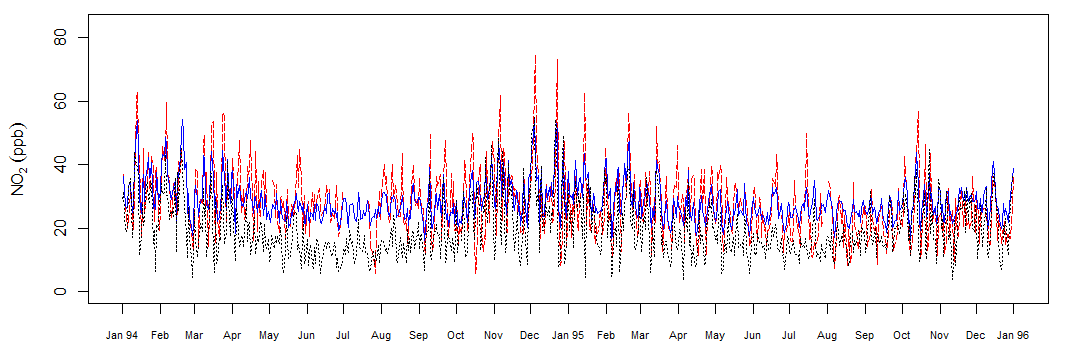}
	\caption[Observed EPA, SCARR fitted, and CMAQ estimated \notwo concentration at New Haven, CT.]{Observed EPA (red), SCARR fitted (blue), and CMAQ estimated (black dotted) \notwo concentration at New Haven, CT.}
	\label{fig:Paper2fittedNewHavenCMAQ}
	\end{subfigure}
	\caption{Comparison of the observed EPA, SCARR model fit, and CMAQ model estimate.}
\end{sidewaysfigure}

The MLE of $\mu_A$ was not significantly different from zero, and was therefore removed from the model. The estimates (and their standard errors) for the remaining parameters in the final model were $\sigma_Z = 22.53~(0.559), \psi_A= 0.593 ~(0.033), \sigma_A=30.32 ~(1.868),$ and $\beta_c=0.713 ~(0.013)$. Figure \ref{fig:Paper2at} shows the smoothed estimate of $A(t)$, calculated using the Kalman filter, along with its 95\% confidence interval. Figure \ref{fig:Paper2fittedNewHavenCI} shows the observed EPA and SCARR fitted concentration of \notwo for New Haven, along with the 95\% confidence interval, while Figure \ref{fig:Paper2fittedNewHavenCMAQ} gives a comparison between the observed EPA, SCARR model predictions, and CMAQ model estimates of \notwo concentration at this site. While the CMAQ model clearly underestimates \notwo concentrations in the summer, SCARR model estimates more closely follow the observed concentrations. Similar plots for the remaining five EPA sites are given in Figures S1--S4 in the Supplement. 

\setkeys{Gin}{width=1\textwidth}

\setkeys{Gin}{width=1\textwidth}

\begin{figure}[!]
	\centering
		\includegraphics{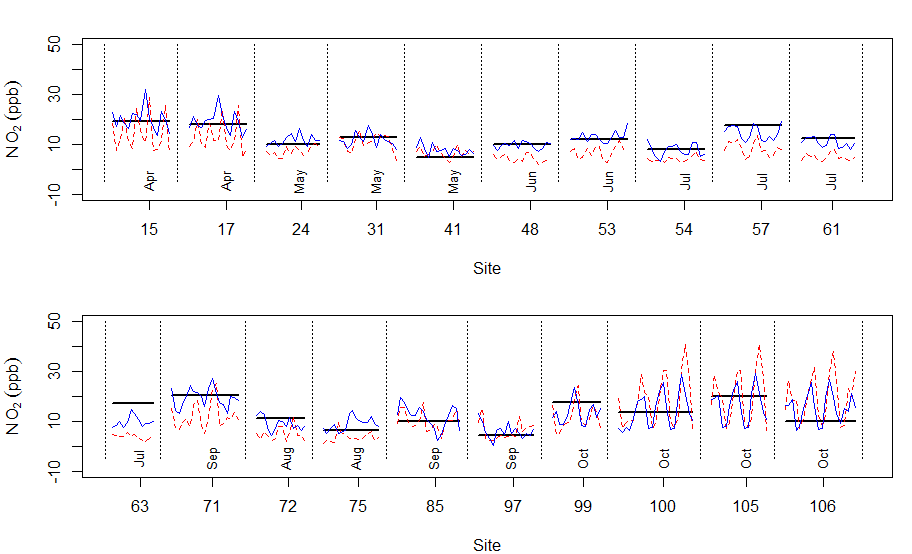}
	\caption[Observed \acidns, SCARR predicted, and CMAQ estimated \notwo concentration at 20 randomly selected \acid sites.]{Observed \acid (black), SCARR predicted (blue), and CMAQ estimated (red) \notwo concentration at 20 randomly selected \acid sites.}
	\label{fig:Paper2ValidationAcid}
\end{figure}

Table \ref{tab:Paper2corr} gives a comparison of the coefficient of correlation and mean squared error (MSE) between the EPA data and estimates from the SCARR and CMAQ models. For all sites except Tolland, SCARR model provides a better fit than CMAQ, with a marked improvement observed at New Haven and Springfield.  Figure \ref{fig:Paper2ValidationAcid} shows a comparison for the observed \notwo concentration with the SCARR and CMAQ model predictions at 20 randomly selected \acid sites. The black horizontal line is the mean \notwo recorded over the 10-14 day period at each site, while the blue and red lines show the daily SCARR and CMAQ model predictions, respectively, for those sites for the same time duration. It also shows the approximate month in which the data were collected. SCARR model predictions are generally closer to the observed \notwo concentrations and provide a much smaller MSPE (36.30) as compared to CMAQ estimates (75.83). A similar plot for all 122 \acid sites is given in Figure S5 in the Supplement.

\begin{table}[h!t]
\caption{Comparing the coefficient of correlation ($r$) and mean squared error (MSE) between EPA and (i) SCARR predictions and (ii) CMAQ predictions.} \label{tab:Paper2corr}
\small
\begin{center}
\begin{tabular}{l c c c c}
\toprule \toprule
 & \multicolumn{2}{c}{\bf{\emph{r}}} & \multicolumn{2}{c}{$\mathbf{MSE}$} \\
\cmidrule(r){2-3} \cmidrule(r){4-5}
 \bf  Site                        &  \bf SCARR   & \bf CMAQ &  \bf SCARR   & \bf CMAQ \\
\midrule

Bridgeport  &   0.700  & 0.693 &   60.62  & 88.68  \\
Chicopee     &  0.789 & 0.749  &  45.00 & 88.59  \\
E. Hartford      & 0.797   &0.763  & 37.05   & 46.78  \\
New Haven  &  0.687 &  0.658 &  55.83 &  123.8 \\
Springfield   & 0.766  &0.681  & 64.36  & 197.4  \\
Tolland        & 0.726  & 0.819  & 21.38  & 18.91  \\

\bottomrule
\end{tabular} 
\end{center}
\caption*{MSE$_s = \sum^{m_s}_{t=1}\frac{(\widehat{Y}(s,t) - Y_3(s,t))^2 }{m_s}$ }
\end{table}

\setkeys{Gin}{width=1\textwidth}


A plot of the standardized residuals against the fitted values from the model as well as a $QQ$ plot of the standardized residuals (not shown) did not reveal any violations of the assumption of normality of the errors. An autocorrelation plot of the residuals over various temporal lags also suggested that the AR(1) assumption for $A(t)$ was justified. 


\section{\notwo Prediction for Connecticut, 1994-1995} \label{Paper2CTpred}

\begin{sidewaysfigure}[!]
	\centering
		\includegraphics{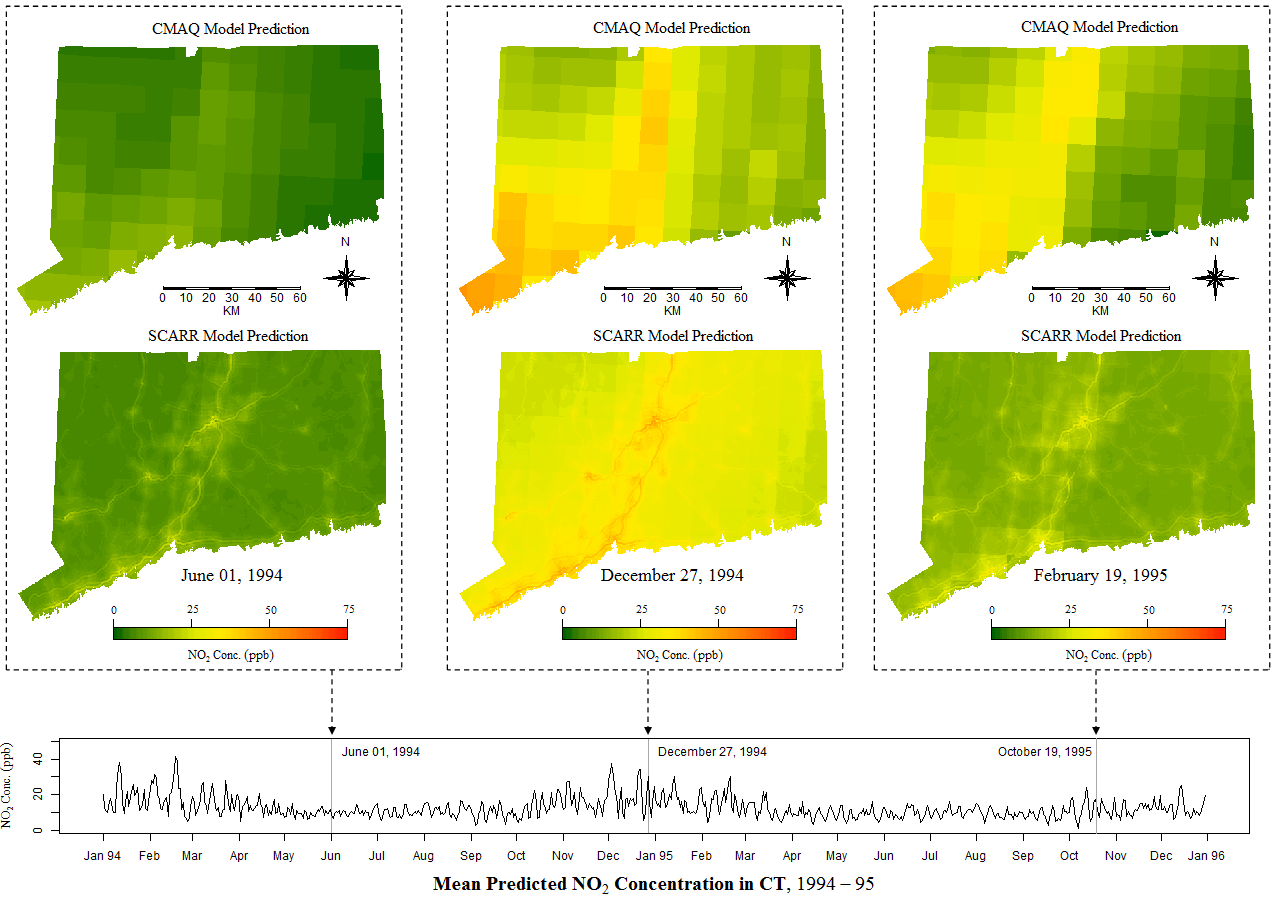}
	\caption{Comparison of \notwo predictions from CMAQ and SCARR models for CT, 1994-1995.}
	\label{fig:Paper2SCARRpred}
\end{sidewaysfigure}

The final model was used to predict the daily concentration of \notwo for the state of \mbox{Connecticut} in 1994 and 1995 over a fine grid with pixels of size \mbox{300 x 300 m.} A raster with 300 m square pixels was created that covered the entire state of Connecticut, and the latitude and longitude of the centroid of each pixel was extracted. For each of the 143050 centroids, the corresponding CMAQ pixels were identified, and the covariates $\ddot{Y}\!_{1}(\s,t)$ and $\widetilde{C}(\s,t)$ were \mbox{calculated} at each location, as detailed in Sections \ref{Paper2Variables} and \ref{Paper2StepIIModel}. To predict a calibrated concentration at a new location $\s'$, the vectors $\mathbf{\widetilde{C}}(t)$ and $\mathbf{\ddot{Y}}\!_{1}(t)$ from equation \ref{eqn:Paper2obsExVec}  are augmented to include $\widetilde{C}(\s',t)$ and $\ddot{Y}\!_{1}(\s',t)$, while the EPA observation for this location $Y_{3}(\s',t)$ is treated as missing. Then, equation \ref{eqn:Paper2obsExVec} is rewritten as
\begin{equation} \label{eqn:Paper2Fit}
\mathbf{Y}_{3}(t) = A(t) \mathbb{1}_7+ \beta_c\mathbf{\widetilde{C}}(t) + \hat{\gamma}\mathbf{\ddot{Y}}\!_{1}(t) + \sigma_zZ(t)\mathbf{I}_7 \hspace{0.5in} t = 1, 2, \ldots, 730,
\end{equation}
and, \small
\begin{equation*} 
E\big(Y_{3}(\s',t)| \mathbf{Y}_{3}(1), \ldots, \mathbf{Y}_{3}(t)\big) = E\big(A(\s',t)| \mathbf{Y}_{3}(1), \ldots, \mathbf{Y}_{3}(t)\big) + E(\beta_c) \widetilde{C}(\s',t) + \hat{\gamma}\ddot{Y}\!_{1}(\s',t), 
\end{equation*} \normalsize
where $E(\beta_c)$ is given by its MLE and $E\big(A(\s',t)| \mathbf{Y}_{3}(1), \ldots, \mathbf{Y}_{3}(t)\big)$ is evaluated by rerunning the Kalman filter on the augmented data vectors using the MLE of the model parameters estimated earlier. The predicted values for each pixel for each day were reassigned to the original raster grid to create 729 raster images. Prediction maps of the daily concentration of \notwo for Connecticut are displayed for a summer, winter and fall day \mbox{(June 1,} 1994;  December 27, 1994; and October 19th, 1995, respectively) in Figure \ref{fig:Paper2SCARRpred} alongside the corresponding maps created using the predictions from the CMAQ model. An animation that shows the change over time in the spatial distribution of ambient \notwo in CT for 1994 and 1995 is available online at ``https://ogilani.shinyapps.io/CTNO2''. 

\setkeys{Gin}{width=\textheight}

\section{Discussion} \label{Paper2discussion}

In the first step of the model, we spatially calibrated and refined the CMAQ \notwo estimates using observed concentrations available from the \acid data, borrowing spatial information from various local covariates while controlling for time in the model. However, while the resultant model improved the spatial resolution of the CMAQ estimates, the calibration was done over very few time points, which were aggregated over long durations: between 10-14 days. Therefore, it didn't provide effective temporal calibration of the daily variability in the CMAQ data. Given the availability of another source of observed data on the concentration of  \notwo (EPA data) that is temporally dense, in the second step we developed a dynamic spatiotemporal model to use the EPA data for temporal calibration of the spatially refined CMAQ estimates from the first step.

The final model from Step I includes population density, total traffic volume buffers up to 2 km, ``forest 0-2 km'' buffer for land use type, and a trigonometric function of date. The model explains about 76\% of the variability in the observed data, with a leave-one-out cross validation PRESS statistic (936) close to the residual sum of squares (752), suggesting it does a decent job at predicting the concentration of \notwo at a new location. The coefficient for total traffic volume buffer 0.5-1 km was less than zero, suggesting that traffic volume within that buffer reduces the concentration of NO$_2$, which is unexpected. However, the estimate is not statistically significant, and was included in the final model to allow the next outer buffer ring, 1-2 km, to be included, which was statistically significant. Similar ``U shaped" estimates for the dispersion function of NO$_2$ have been observed in previous studies \citep{holford2010}, and may be explained by the complex process that produces NO$_2$ from nitrogen oxide (NO) and ozone (O$_3$) in the atmosphere. NO$_2$ is not produced directly by combustion in automobiles, and elevated levels of NO$_2$ are often observed at farther distances, perhaps reflecting the time needed to convert NO to NO$_2$. The estimated trigonometric function (Figure \ref{fig:Paper2Season}) shows two peaks within a 12 month cycle - a higher peak in the winter and a lower one in the summer. While the fit of the curve seems appropriate for these data, comparing the time trend of \notwo for the EPA data suggests that the summer peak estimated by the trigonometric function might be artificially high in the \acid data. However, an advantage of the two stage modeling strategy is that this possible over estimation in the summer months in Step I is countered during the temporal calibration in Step II, as seen by a dip in the mean trend of the estimate of $A(t)$ (Figure \ref{fig:Paper2at}) during the corresponding summer period. 

The MLE for the parameter $\mu_A$ in Step II of the model was not statistically significantly different from zero. Given the presence of $\widetilde{C}(\s,t)$ in the model, this was to be expected as $\widetilde{C}(\s,t)$  captures the mean spatial additive calibration bias for the CMAQ data. The parameter $\beta_c$ controls the influence of $\widetilde{C}(\s,t)$ in Step II. The MLE for $\beta_c$ was 0.713, which suggests a slight mitigation of the spatial additive calibration bias $\widetilde{C}(\s,t)$, estimated in Step I of the model, when evaluating the temporal evolution of the additive calibration bias $A(t)$ in the second step.  

As discussed in \citet{gilani2016}, the CMAQ model does a reasonable job of predicting the ambient \notwo concentrations in the winter months when the concentrations are generally high. But it does not provide very accurate predictions during the low concentration periods in the summer, when it generally underpredicts the true concentration. As seen in \mbox{Figure \ref{fig:Paper2fittedNewHavenCMAQ},} the SCARR model improves the prediction during the summer months and, overall, appears to provide a better prediction as compared to the CMAQ model. The coefficients of correlation  comparing the EPA observations with predictions from the SCARR and CMAQ model, and the empirical mean squared errors (MSE) for these two models (Table \ref{tab:Paper2corr}), show that the SCARR model provides better predictions at five of the six sites, with a remarkable improvement in the prediction for New Haven and Springfield. The CMAQ model, on the other hand, appears to provide better predictions at Tolland. However, as mentioned in \S \ref{Paper2StepIIResults}, the site at Tolland was missing EPA data for the summer of 1994, and the correlation and MSE for this site reflect the performance of the two models only for the winter period, during which time the CMAQ model generally performs well. For almost all of the \acid sites, predictions from the SCARR model more accurately reflect the truth than the CMAQ model (Figures \ref{fig:Paper2ValidationAcid} and S5). The empirical MSPE for the SCARR model predictions (36.30) at the \acid sites was much smaller than the MSPE for the CMAQ model prediction (75.83). 

There are a few limitations to the modeling strategy presented here. The model assumes that the \notwo observations recorded at the \acid and EPA sites accurately reflect the true ambient concentrations. However, qualitatively that might not necessarily be the case as different monitoring equipments record \notwo concentrations at varying levels of accuracy. Additionally, the EPA monitors are typically placed in high concentration locations near major roadways. However, the two step modeling strategy may actually help in balancing this effect by also including observations from the \acid data, whose locations were sampled independent of pollutant concentrations and were therefore more evenly distributed between high and low concentration areas.  Another limitation is due to the fact that Step II of the model includes variables that were estimated in Step I ($\widetilde{C}$), and the resulting estimates of the model errors do not capture the additional uncertainty of including these estimated variables, leading to somewhat conservative prediction errors. Methods developed for accounting for measurement errors in models can be applied to this model to provide more accurate error estimates \citep{carroll2006}. Given the Markov property of Kalman filters, a Bayesian approach can also be utilized to explicitly account for the uncertainty in the estimated variables included in the model. However, prediction of \notwo on a fine spatiotemporal resolution, with 143,050 spatial locations and 730 time points, using a Bayesian approach can impose a \mbox{significant computational burden.} 

The \notwo prediction maps for CT for 1994 and 1995 (Figure \ref{fig:Paper2SCARRpred}) show a clear improvement in the spatial resolution as compared to the estimates provided by the CMAQ model. The effect of local covariates is evident in the finer spatial resolution map, where the contribution of traffic on major highways to ambient \notwo concentration stands out. These maps provide more accurate estimates for points within the CMAQ pixel. For example, the concentration of \notwo appears rather high at all points in the pan-handle of CT (south-west corner of the map) in the CMAQ model estimates, whereas the SCARR model predictions show that the concentration is high primarily along the major highway (Interstate 95), while locations away from the highway have a significantly lower concentration. These maps with more accurate estimates at the centroid and finer spatial resolution can be very useful in assigning mean daily exposure to \notwo for participants in epidemiologic studies, which is usually not possible to do using available observed sources of data on the concentration of NO$_2$. Predictions from this model significantly contribute to exposure assessment at a fine spatial and temporal resolution for CT in 1994 and 1995.

\section*{Acknowledgement}

The authors thank Dr. Lance Waller for useful feedback on the manuscript. This research was partially funded by grant R01ES017416 from the National Institutes of Health.

\bibliographystyle{apalike}
\bibliography{biblio}

\end{document}


\setkeys{Gin}{width=1.0\textwidth}  

\title{Spatiotemporal Calibration of Atmospheric Nitrogen Dioxide Concentration Estimates From an Air Quality Model for Connecticut \\ \vspace{0.2in} Supplementary Material} 


\author{Owais Gilani$^{a}$\thanks{Corresponding author: Tel: +1 570-577-1391; Email: owais.gilani@bucknell.edu}, Lisa A. McKay$^b$, Timothy G. Gregoire$^c$, \\
 Yongtao Guan$^d$,   Brian P. Leaderer$^b$, Theodore R. Holford$^b$ \vspace{0.1in} \\ 
\emph{\small $^a$Bucknell University, Lewisburg, PA 17837, USA} \\ \emph{\small $^b$Yale School of Public Health, Yale University, New Haven, CT 06510, USA} \\ \emph{\small $^c$Yale School of Forestry \& Environmental Studies, New Haven, CT 06511, USA}  \\  \emph{\small $^d$School of Business Administration, University of Miami, FL 33124, USA } }

%
%
%
%
%
%
%
%

\maketitle

Figure \ref{fig:Paper2fitted6CI1} shows the observed EPA (red) and the SCARR model fitted (blue) concentrations of \notwo along with the 95\% confidence intervals (black dotted) for Bridgeport, Hartford and New Haven, Connecticut (CT) for 1994 and 1995, while Figure \ref{fig:Paper2fitted6CI2} provides similar plots for Tolland, CT, and Springfield and Chicopee, Massachusetts (MA).

Figure \ref{fig:Paper2fitted6CMAQ1} gives a comparison between the observed EPA (red), SCARR model predictions (blue), and CMAQ model estimates (black dotted) of \notwo concentration at Bridgeport, Hartford, and New Haven, CT, while Figure \ref{fig:Paper2fitted6CMAQ2} gives similar comparisons at Tolland, CT, and Springfield and Chicopee, MA. The SCARR model predictions more closely follow the observed EPA concentrations as compared to the CMAQ model estimates, particularly in the summer months. 

Figure \ref{fig:Paper2ValidationAcidAll} shows a comparison of the observed \notwo concentration with the SCARR and CMAQ model predictions at 122 \acid sites. The black horizontal line is the mean \notwo recorded over the 10-14 day period at each site, while the blue and red lines show the daily SCARR and CMAQ model predictions, respectively, for those sites for the same time duration. It also shows the approximate month in which the data were collected. The empirical mean squared prediction error (MSPE) for the SCARR and CMAQ model predictions at these sites were 36.30 and 75.83 respectively.

\renewcommand\thefigure{S\arabic{figure}} 
\setkeys{Gin}{height=0.8\textwidth, width=\textheight}
\setcounter{figure}{0}

\begin{sidewaysfigure}[!]
	\centering
		\includegraphics{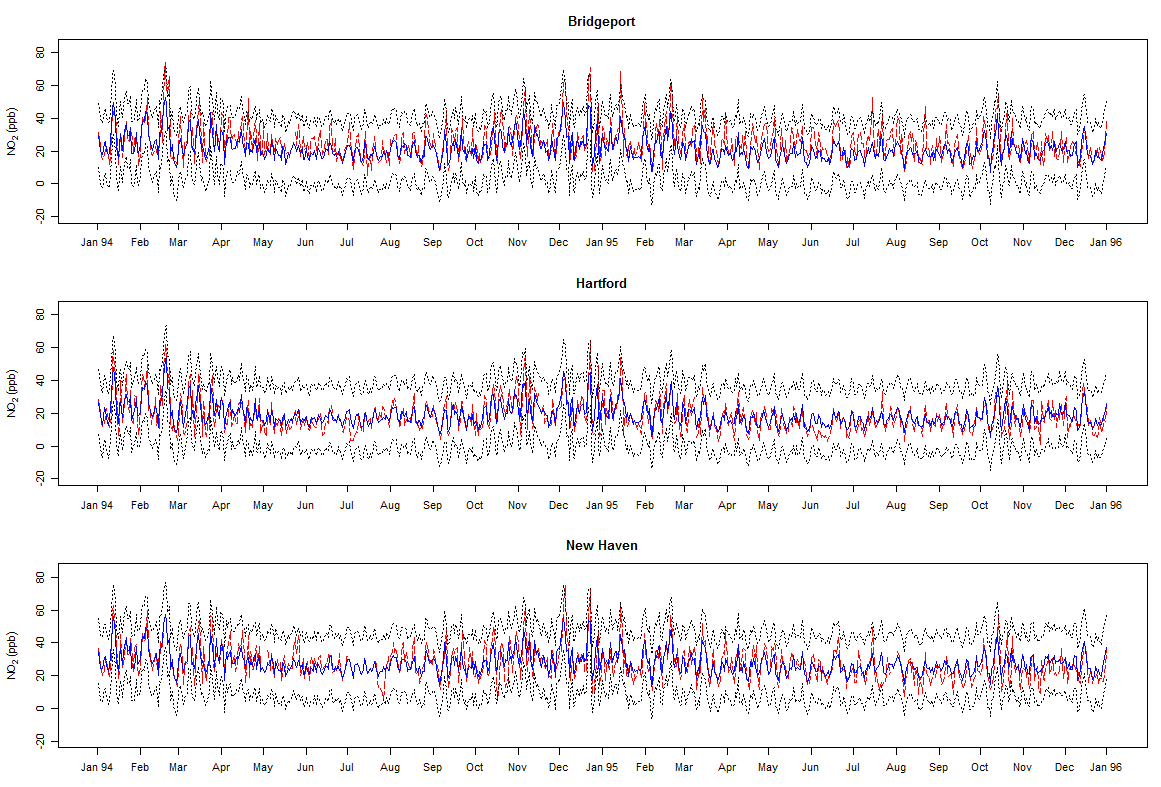}
	\caption[Observed EPA and SCARR fitted \notwo concentration at Bridgeport, East Hartford and New Haven, with 95\% confidence intervals.]{Observed EPA (red) and SCARR fitted (blue) \notwo concentration at Bridgeport, East Hartford and New Haven, with 95\% confidence intervals (black dotted).}
	\label{fig:Paper2fitted6CI1}
\end{sidewaysfigure} 

\setkeys{Gin}{height=0.8\textwidth, width=\textheight}
\begin{sidewaysfigure}[!]
	\centering
		\includegraphics{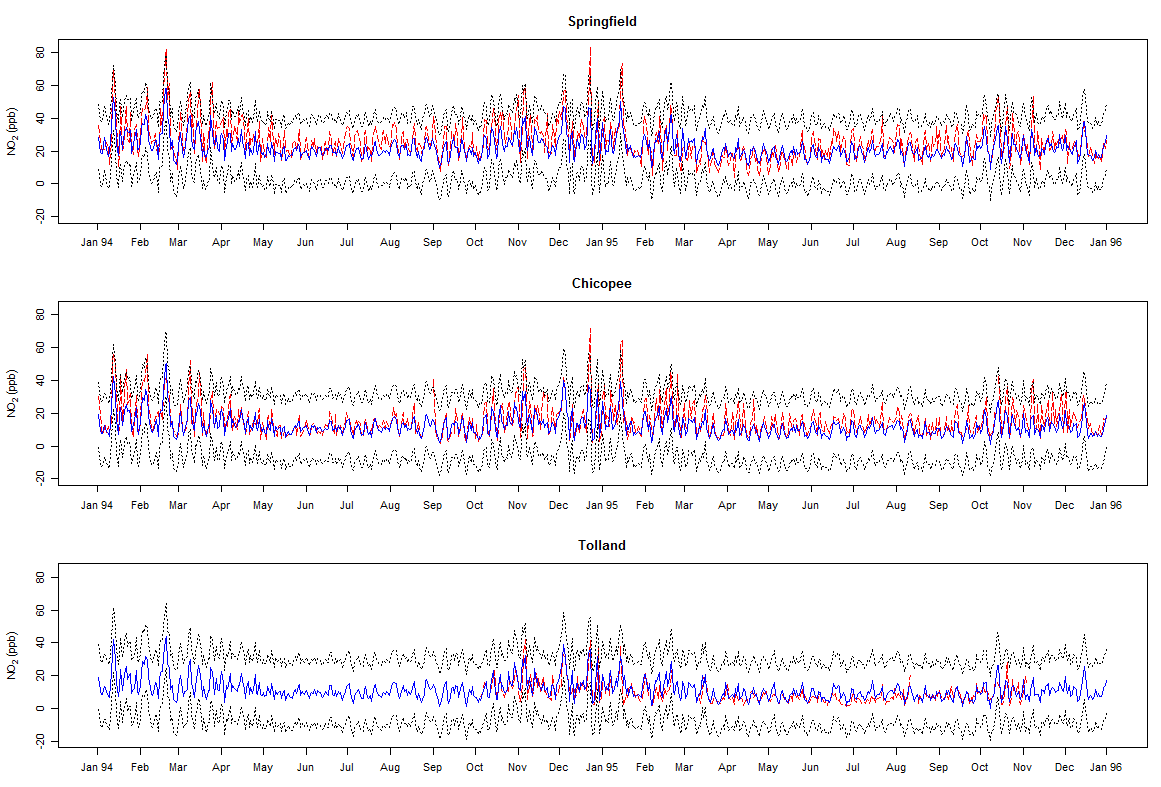}
	\caption[Observed EPA and SCARR fitted \notwo concentration at Tolland, Chicopee, and Springfield, with 95\% confidence intervals.]{Observed EPA (red) and SCARR fitted (blue) \notwo concentration at Tolland, Chicopee, and Springfield, with 95\% confidence intervals (black dotted).}
	\label{fig:Paper2fitted6CI2}
\end{sidewaysfigure} 

\begin{sidewaysfigure}[!]
	\centering
		\includegraphics{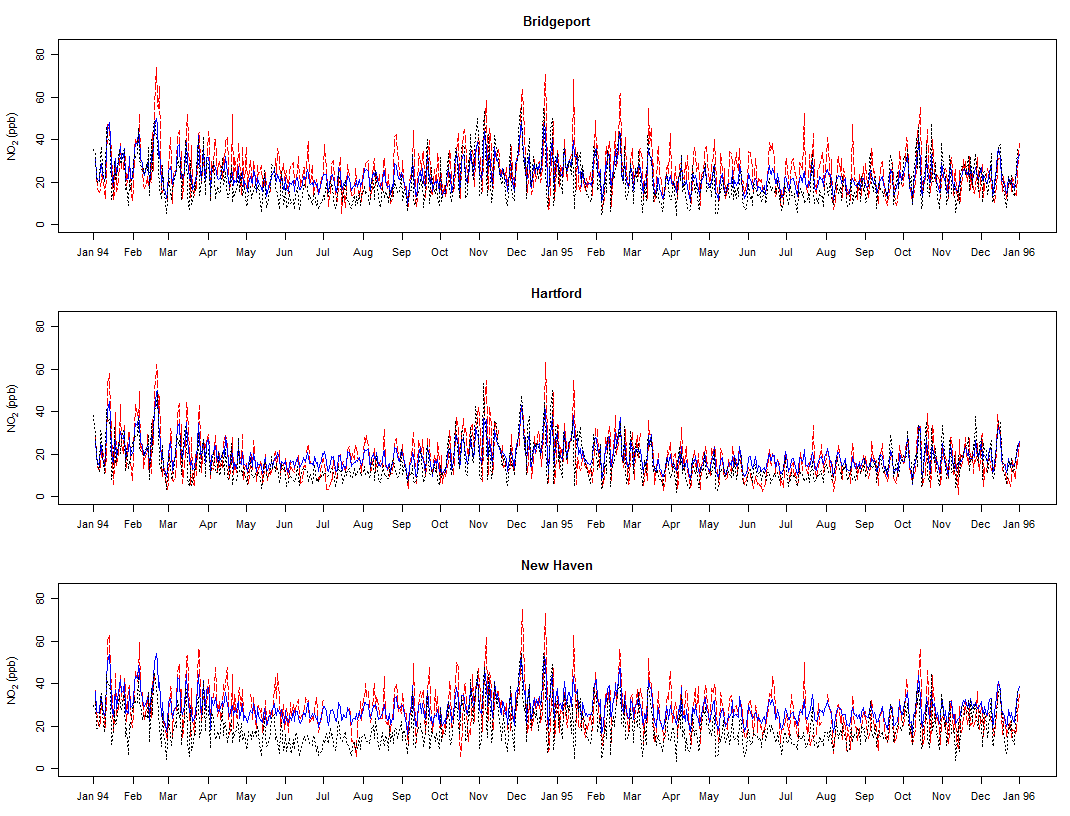}
	\caption[Observed EPA, SCARR fitted, and CMAQ estimated \notwo concentration at Bridgeport, E. Hartford and New Haven.]{Observed EPA (red), SCARR fitted (blue), and CMAQ estimated (black dotted) \notwo concentration at Bridgeport, E. Hartford and New Haven.}
	\label{fig:Paper2fitted6CMAQ1}
\end{sidewaysfigure} 

\begin{sidewaysfigure}[!]
	\centering
		\includegraphics{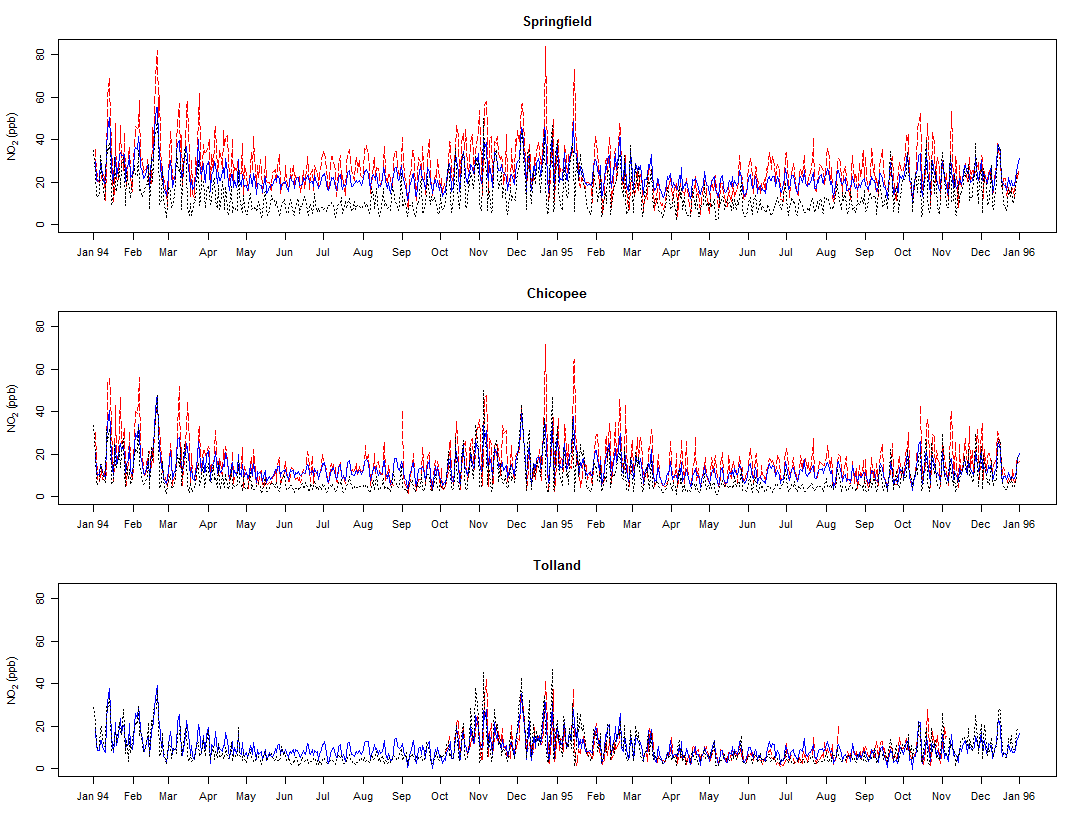}
	\caption[Observed EPA, SCARR fitted, and CMAQ estimated \notwo concentration at Tolland, Chicopee and Springfield.]{Observed EPA (red), SCARR fitted (blue), and CMAQ estimated (black dotted) \notwo concentration at Tolland, Chicopee and Springfield.}
	\label{fig:Paper2fitted6CMAQ2}
\end{sidewaysfigure} 

\setkeys{Gin}{width=1\textwidth}
\begin{sidewaysfigure}[!]
	\centering
		\includegraphics{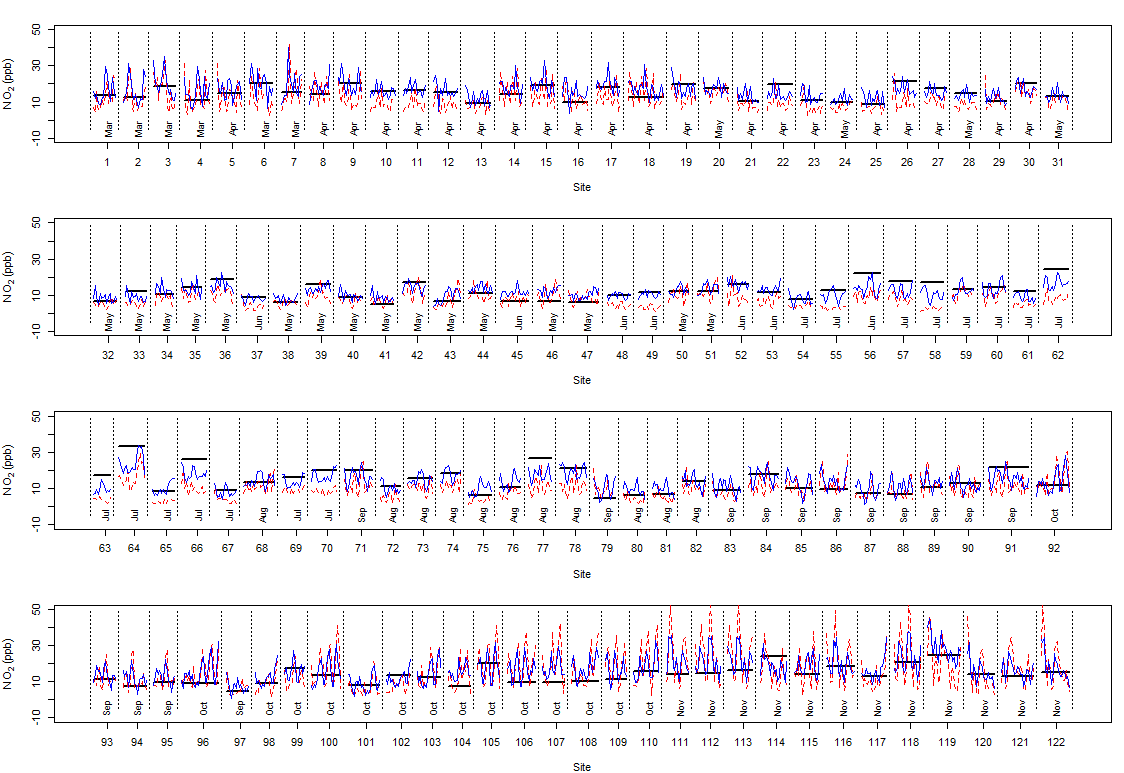}
	\caption[Observed \acidns, SCARR predicted, and CMAQ estimated \notwo concentration at 122 \acid sites.]{Observed \acid (black), SCARR predicted (blue), and CMAQ estimated (red) \notwo concentration at 122 \acid sites.}
	\label{fig:Paper2ValidationAcidAll}
\end{sidewaysfigure}